\documentclass[sigconf,screen,authorversion]{acmart}

\usepackage{float}
\usepackage{booktabs}
\usepackage{multirow}

\usepackage{tikz}
\usetikzlibrary{arrows.meta, positioning}
\AtBeginDocument{%
  }

\copyrightyear{2026}
\acmYear{2026}
\setcopyright{cc}
\setcctype{by}
\acmConference[LAK 2026]{LAK26: 16th International Learning Analytics and Knowledge Conference}{April 27-May 01, 2026}{Bergen, Norway}
\acmBooktitle{LAK26: 16th International Learning Analytics and Knowledge Conference (LAK 2026), April 27-May 01, 2026, Bergen, Norway}

\acmDOI{10.1145/3785022.3785128}
\acmISBN{979-8-4007-2066-6/2026/04}

\begin{document}

\title[Session-by-Session Analytics of Teacher Interventions in K-12 Classrooms]{Sticky Help, Bounded Effects: Session-by-Session Analytics of Teacher Interventions in K-12 Classrooms}

\author{Qiao Jin}
\orcid{0000-0001-5493-1343}
\affiliation{
    \institution{North Carolina State University}
    \city{Raleigh, NC}
    \country{USA}
}
\email{qjin4@ncsu.edu}

\author{Conrad Borchers}
\orcid{0000-0003-3437-8979}
\affiliation{
    \institution{Carnegie Mellon University}
    \city{Pittsburgh, PA}
    \country{USA}
}
\email{cborcher@cs.cmu.edu}

\author{Ashish Gurung}
\orcid{0000-0001-7003-1476}
\affiliation{
    \institution{Carnegie Mellon University}
    \city{Pittsburgh, PA}
    \country{USA}
}
\email{agurung@andrew.cmu.edu}

\author{Sean Jackson}
\orcid{0009-0008-6572-7071}
\affiliation{
    \institution{Carnegie Mellon University}
    \city{Pittsburgh, PA}
    \country{USA}
}
\email{seanjack@andrew.cmu.edu}

\author{Sameeksha Agarwal}
\orcid{0009-0007-9963-7288}
\affiliation{
    \institution{Carnegie Mellon University}
    \city{Pittsburgh, PA}
    \country{USA}
}
\email{sameeksa@andrew.cmu.edu}

\author{Cancan Wang}
\orcid{0009-0006-2226-7242}
\affiliation{
    \institution{Carnegie Mellon University}
    \city{Pittsburgh, PA}
    \country{USA}
}
\email{cancanw@andrew.cmu.edu}

\author{YiChen Yu}
\orcid{0009-0001-0175-3253}
\affiliation{
    \institution{North Carolina State University}
    \city{Raleigh, NC}
    \country{USA}
}
\email{yyu55@ncsu.edu}

\author{Pragati Maheshwary}
\orcid{0009-0002-2216-5394}
\affiliation{
    \institution{Carnegie Mellon University}
    \city{Pittsburgh, PA}
    \country{USA}
}
\email{pragati2@andrew.cmu.edu} 

\author{Vincent Aleven}
\orcid{0000-0002-1581-6657}
\affiliation{
    \institution{Carnegie Mellon University}
    \city{Pittsburgh, PA}
    \country{USA}
}
\email{aleven@cs.cmu.edu}

\renewcommand{\shortauthors}{Jin et al.}

\begin{abstract}

Teachers' in-the-moment support is a limited resource in technology-supported classrooms, and teachers must decide whom to help and when during ongoing student work. However, less is known about how students' prior help history (whether they were helped earlier) and their engagement states (e.g., idle, struggle) shape teachers' decisions, and whether observed learning benefits associated with teacher help extend beyond the current class session. To address these questions, we first conducted interviews with nine K–12 mathematics teachers to identify candidate decision factors for teacher help. We then analyzed 1.4 million student–system interactions from 339 students across 14 classes in the MATHia intelligent tutoring system by linking teacher-logged help events with fine-grained engagement states. Mixed-effects models show that students who received help earlier were more likely to receive additional help later, even after accounting for current engagement state. Cross-lagged panel analyses further show that teacher help recurred across sessions, whereas idle behavior did not receive sustained attention over time. Finally, help coincided with immediate learning within sessions, but did not predict skill acquisition in later sessions, as estimated by additive factor modeling. These findings suggest that teacher help is ``sticky'' in that it recurs for previously supported students, while its measurable learning benefits in our data are largely session-bound. 
We discuss implications for designing real-time analytics that track attention coverage and highlight under-visited students to support a more equitable and effective allocation of teacher attention.
\end{abstract}



\maketitle

\section{Introduction}

Teachers’ in-the-moment support can shape students’ engagement and immediate progress during technology-enhanced classrooms~\cite{holstein_designing_2022, martinez-maldonado_handheld_2019, possaghi_integrating_2025}. However, because teachers must make rapid decisions under limited time while monitoring many students, understanding what drives their intervention decisions has been a central challenge for learning analytics~\cite{slotta_orchestrating_nodate,aleven_developing_nodate}. In classrooms, teachers often make decisions based on partial signals from students’ ongoing work (e.g., help-seeking), because students’ underlying effort is not directly observable~\cite{gurung2021examining}. To support this real-time orchestration, Intelligent Tutoring System (ITS) log data are commonly translated into real-time analytics that summarize student progress and participation, helping teachers decide whom to help and when~\cite{holstein_intelligent_2017, jin2025who}. Prior work emphasizes that such decision support is most useful when indicators are concise and interpretable, avoid disruptive alerts, and preserve teacher autonomy in how analytics are acted upon~\cite{holstein_co-designing_2019, an_ta_2020}. Teacher dashboards built for ITS classrooms (e.g., Lumilo~\cite{holstein2018student}, MTDashboard~\cite{martinez2013mtclassroom}, Pair-Up~\cite{yang2023pair}, and DREAM~\cite{tissenbaum2016real}) operationalize these ideas by presenting fine-grained interaction evidence, such as hint requests, error patterns, and step-level mastery estimates, in forms intended to align with teachers’ instructional decision-making needs~\cite{aleven_developing_nodate}.
Beyond tool design, research examines how teachers use analytics in class and how dashboards shape feedback. For example, classroom studies show that teacher dashboards can change how feedback is distributed across students, including making feedback allocation more even across different performance levels~\cite{knoop2023equalizing}. Observational and case-based work suggests substantial variation in how teachers consult dashboards, interpret indicators, and translate them into concrete actions, shaped by teachers' routines and characteristics~\cite{molenaar2019designing,campen2020teachers,van2021teacher}.

At the same time, most orchestration and learning analytics studies model students’ current states (e.g., idleness, repeated struggle, off-task behavior)~\cite{baker_modeling_2007, beck2013wheel, booth_engagement_2023} and report that teacher interventions can reduce momentary disengagement~\cite{karumbaiah2023spatiotemporal,jin2025who}. However, classroom interventions are inherently temporal, both within and across sessions: a student’s history of receiving help may shape teachers’ future decisions, for instance, by avoiding over-support or ensuring fairness. Despite calls for more longitudinal approaches to modeling teacher-student interactions~\cite{hamaker_critique_2015, karumbaiah2023spatiotemporal, gurung2025starting}, factors of help history and cross-session effects remain underexplored at scale in authentic classrooms.

Our research asks how teachers decide whether to visit and help a student during technology-supported classroom work, and whether these help visits yield benefits that persist across classwork sessions. We use a mixed-methods approach with two components: interviews with teachers and large-scale log analysis combining student ITS interaction data with teacher-entered dashboard records of visits. We first interviewed nine middle-school math teachers about why and how they provide in-the-moment support, and what factors guide decisions about when to initiate visits. Teachers consistently highlighted students’ prior help history and engagement state (e.g., idle and struggle) as key drivers of these decisions, alongside contextual considerations. These insights motivate our quantitative models, which treat help history and engagement state as key predictors of teacher help. 
In the quantitative analysis, in order to test whether students' help history and momentary engagement states predict teachers' help visits at scale, and to assess whether teacher visits are associated with improvements that persist into later classwork sessions, we follow prior practice~\cite{gurung2025starting} to align the analysis window with when teachers can act and when help is logged. Accordingly, we test these factors on a large-scale classroom log dataset from MATHia (1.4M student–system interactions) at the session level (defined as a contiguous in-class period inferred from logs; homework and other out-of-class activity are excluded). Within this session-based frame, we also include delayed start (time before initiating work) as a stability-sensitive indicator of motivation: prior work shows delayed start displays trait-like stability across sessions and predicts outcomes above and beyond teacher behavior~\cite{gurung2025starting}. We are interested in whether some disengagement patterns are stable and, if so, whether teacher visits lead to prospective improvements across sessions or whether teachers primarily revisit the same students in a reactive, sticky pattern. Relatedly, analytics that quantify the effectiveness of visits and revisits across sessions could help refine dashboards that currently focus on momentary disengagement and better support competing teacher priorities.

Building on the interview accounts and our session-based log analyses, we ask:

\begin{itemize}
    \item \textbf{(RQ1)} Do help history and student states (idle, struggle) affect which students receive teacher help?
    \item \textbf{(RQ2)} Across sessions and after accounting for student disengagement stability, does prior teacher help predict future student states, or, alternatively, do prior student states predict future teacher help?
    \item \textbf{(RQ3)} Do teachers' help relate to improvements in students’ learning in future sessions?
\end{itemize}

To examine the likelihood of teacher help at the student level \textbf{(RQ1)}, we utilize generalized linear mixed models (GLMM) \cite{bates_fitting_2015}, incorporating factors such as help history, student states, delayed start indicating struggle and idle, session type, and random effects for class and student nesting. To investigate potential temporal effects of help history and student engagement, we employ cross-lagged panel models (CLPM)~\cite{selig2012autoregressive} to assess directional influences between teacher help and students' states \textbf{(RQ2)}, and using additive factor model (AFM)-based indicators~\cite{effenberger2020exploration} to test whether help predicts later skill acquisition \textbf{(RQ3)}. 

This study contributes to learning analytics by (1) characterizing teacher-reported goals, support types, and decision drivers for providing help in technology-enhanced classrooms through teacher interviews; (2) providing log-based evidence that prior teacher help increases the odds of subsequent help, with implications for session-by-session analytics and dashboard designs that aim to support more equitable allocation of teacher attention; (3) offering evidence that the learning benefits associated with help are largely session-bound in our data, with limited evidence of carryover to later sessions; and (4) informing teacher orchestration practice and analytics tool design to reduce concentrated revisits and surface cross-session disengagement signals by synthesizing teacher-reported goals with empirical patterns in the data.

\section{Teacher Interviews on In-Class Help Events in ITS-Supported Classrooms}
We began with an interview study to understand how teachers working with an ITS describe their role and the choices they make during class. The aim was to map teachers’ stated goals, the types of support they provide during \emph{in-class help events} (``visits''), and the factors teachers report using to guide these decisions. This qualitative step grounds our subsequent log analyses and helps interpret cross-session patterns.

\subsection{Participants and Settings}
The study was approved by the university’s Institutional Review Board (IRB). Through partner schools and social media recruitment, we invited nine middle‑school mathematics teachers (grades 6–9) from both public and private schools to participate in this study. Seven teachers worked in public schools and two in a private school. Six participants identified as female and three as male.
Participants had diverse teaching experience, from early-career (around 5 years, n=2), mid-career (10 to 15 years, n=3), to veteran (over 16 years, n=4), and typical class sizes ranged from 10 to 36 students. All participants reported experiences of using a classroom setup where students directly interact with one or more ITS (e.g., MATHia,\footnote{\url{https://www.carnegielearning.com/solutions/math/mathia}} Khan Academy,\footnote{\url{https://www.khanacademy.org/}} ASSISTments\footnote{\url{https://www.assistments.org/}}) for their math class. 
We stopped recruiting once we reached data saturation, meaning that additional interviews yielded no new themes and teachers’ responses began to converge. Each participating teacher received a \$30 gift card for their time. All interviews were audio-recorded and transcribed for analysis.

\subsection{Data Collection and Analysis}
\label{datacollection}
We conducted 60-minute, semi-structured videoconference interviews with participating teachers. The protocol aimed to serve two purposes: first, to document how teachers describe their in‑class intervention decisions during ITS‑supported lessons; second, to use the tech‑exposure segment to establish a shared vocabulary for engagement indicators so that teachers can reflect on which signals are actionable, which they can ignore, and what is lacking for practice. 
At the start of the interview, we obtained consent and verified demographics from a pre-session questionnaire. We then collected class-level context (e.g., grade level, class size, ITS usage) and asked about current ITS intervention practices, intervention goals, and the types of interventions used in class.

To ensure a common reference point for discussion, we shared a video demo of a mixed-reality, teacher-facing dashboard (Lumilo~\cite{holstein_classroom_2018}) and walked teachers through its engagement state indicators. This walkthrough helped align terminology and enabled teachers to comment on which signals they would act on when initiating help visits. We chose Lumilo because it presents a well-defined set of states: idle (no interactions in the software for 2 minutes or more~\cite{holstein2018student}), struggle (students attempting to master a skill three or more times without mastering it despite system support~\cite{beck2013wheel}), misuse (rapid guessing, abusing hints, or gaming the system~\cite{baker_off-task_nodate}), doing well (low recent error rate in current activity, defined as an error rate less than 20\% over the last 10 actions ~\cite{holstein2018student}). These states map naturally to ITS-supported in-class actions and have been studied with K–12 teachers, with evidence that it improves student learning outcomes~\cite{holstein_classroom_2018}. We then asked teachers to recount a recent lesson and explain whom they helped first, what cues drew their attention when deciding whether to initiate a help visit (with or without Lumilo), and how they balanced individual support with whole‑class. We closed by inviting any additional comments about data or factors that would help them decide when and whom to visit.

We used reflexive thematic analysis~\cite{braun2019reflecting} to identify teachers’ reported goals, types of intervention, and recurrent decision factors. Two researchers independently read all transcripts in full, wrote analytic memos, and conducted inductive open coding. The team iteratively refined the coding framework through discussion; weekly peer-debrief meetings with a third researcher were used to resolve disagreements through consensus. We then organized coded segments into candidate themes, reviewed them against disconfirming cases, and refined theme definitions and boundaries. 

\subsection{Findings}
\label{formativefindings}
\subsubsection{Goals of Interventions}
\label{goals}

\textbf{Maximize learning while keeping the lesson moving}

Teachers aimed to maximize learning while keeping students moving forward in their problem-solving practice. They adapted pacing, content, and instructional strategies in real time, particularly when addressing accumulating errors. As P3 noted, ``I break lessons into chunks and check progress mid-way to redirect if errors build up'' (P3), ensuring learning stays on track. P7 added, ``If students lag on early steps, I shift examples to them to clarify before moving on'' (P7). P6 further emphasized, ``Seeing many idle signals means a lesson flaw, so I replan on the spot to keep us moving forward'' (P6).

\textbf{Sustain student engagement}
Teachers restructured activities (e.g., groups, demos) to maintain active participation, closely monitoring signals like work pace to keep students involved. P4 adjusted dynamically to foster activity: ``I switch to group tasks if pace slows, watching who’s engaged'' (P4). Some teachers observed cues to respond to the disengagement: ``I check for idle students and regroup them to boost participation'' (P7) and ``If a student’s work stalls (idle indicator), I demo a step and get them moving again'' (P9).

\textbf{Maintain an inclusive climate}
Teachers fostered safe help-seeking and balanced attention to avoid stigma, creating an inclusive environment. P1 emphasized discretion: ``When I walk, I'm not asking them to lose face... I'm able to just like, work with them individually without drawing a lot of attention to it'' (P1). P5 promoted equity through ``switch seats every two weeks... reorganize groups based on data (from the teacher dashboard) to regroup based on data for fairness'' (P5). P1 described language-inclusive support by pairing students with peers who share the same language.

\textbf{Responses to individual differences}
Teachers customize for language, Individualized Education Program (IEP)/Attention-Deficit
/Hyperactivity Disorder (ADHD), or emotional needs, using scaffolds or deferrals. P1 tailored support for multilingual students: ``Sometimes I will group them together with that support to see if he can explain things in Spanish'' (P1). Teachers noted other individual needs like``some students have like they need redirection... You have to know which ones need that motivation'' (P1) and ``[One student] is more likely to talk with others, so I have to keep reminding him to focus on the task'' (P7). 

\subsubsection{Types of Interventions}
\label{types}

\textbf{Content guidance}
Teachers provided targeted academic support to address misconceptions or stalled progress, such as error correction, examples, and step breakdown. P1 described such scaffolding ``If a student isn’t writing anything, I go over and say, `Okay, here’s where we start.' And then I say, `Now, what would you add to this?''' (P1). P3 offered similar support: ``I break lessons into chunks and check progress mid-way to redirect if errors build up'' (P3). P7 reinforced academic reflection: ``I ask them to explain their steps and choose a better method if needed'' (P7).

\textbf{Motivational and behavioral guidance}
Teachers used motivational and behavioral cues to sustain engagement and support emotional well-being. For example, P1 offered praise when students ``raised their hand, or they asked for help'', walking over to students when they looked frustrated and telling them ``You can do this.'' P5 gave a short break to help students refocus ``if a student seems overwhelmed'' (P5). Similarly, P9 reinforcing positive behavior by walking around ``giving gems to keep them trying, especially if they’re lagging'' (P9). 

\textbf{Tool/technology support}
In some cases, teachers visit students to address technical barriers on operation and system troubleshooting. P1 noted his experience of ``Teaching them how to navigate through a to-do list or navigate through a lesson on their own'' (P1). P3 referred to a troubleshooting case as ``If the system lags, I step in to fix it so they can keep working'' (P3). %

\subsubsection{Factors Influenced Interventions}
\label{factors}

\textbf{Striving for equitable coverage}
Teachers described their continuous effort to achieve the fairest possible distribution of attention in teaching, even in large classes. P1 summarized this goal ``when you have 32 students in a class, you always want education to be extremely, extremely fair—and the current model isn't designed for fairness...'' (P1). 
Proactive planning to avoid blind spots is a common practice, as P5 noted: ``Rearrange groups based on exit tickets or baseline data to ensure no one is left behind'' (P5). P7 emphasized closing the loop with every learner: ``Whenever any student points out an error needing correction, I will (assist)... ensuring everyone understands, ensuring everyone is included in the learning process'' (P7). P9 linked in-class support with after-class follow-up to avoid spending too much time on the same students during the class: ``If you already help these students twice... Probably they have to be dealt with one-on-one later'' (P9).

\textbf{Prioritizing current engagement state}
All teachers agreed that engagement state indicators are useful for deciding when to intervene. Among all the learning indicators we demoed during the interview, \textit{idle} and \textit{misuse} were most often ranked as highest-priority cues that trigger a teacher visit. Although instructors considered positive indicators such as ``doing well'' useful, they seldom triggered action; when they did, it was typically limited to a brief acknowledgment (e.g., a quick check-in or praise). As P9 proposed a tiered handling principle, ``Prioritize idle states as warning signals... followed by inefficient learning, then learning difficulties. ...Of course, high-achieving students (with doing well indicator) require the least concern'' (P9). 
P3 endorsed this prioritization ``Address inactivity first... then focus on learning difficulties (struggle indicator), identifying which step went wrong'' (P3). %

\textbf{Prioritizing peer-led support.}
Most teachers prefer to first guide peer or group collaboration to foster autonomy, gradually expanding support before intervening. As P1 stated: ``70\% of the time they can solve problems by working with teammates... Listen to their explanations before rushing to intervene... If they understand the lesson content, let them help their peers'' (P1). P9 linked this to values cultivation: ``Empathy toward fellow learners is a core competency I want students to develop... Pair children with peers who can offer assistance'' (P9). When misunderstandings persist, intervention is deliberate: ``If summaries contain errors or cognitive biases... teachers guide students to adjust their interpretations through problem-solving practice'' (P1).

\textbf{Contextual observations and individual needs}
Interventions must also incorporate real-time awareness of classroom dynamics, including student posture, facial expressions, pacing, and support needs. P1 describes continuous observation: ``Scanning the classroom to determine which students are performing tasks appropriately or not... I frequently move to pick up on signals'' (P1). P5 emphasizes ambiguity checks and flexible responses: ``If a student raises their hand but looks confused or frowns, I confirm rather than assume they are idle; some days students need breaks due to emotional fluctuations'' (P5). P7 underscores the importance of anticipating needs: ``Seeing students struggle to express themselves or needing extra time, like those with individualized education plans, walking around allows me to spot these signs early'' (P7). P9 adds silent intervention for emotional barriers: ``If a student slumps or hesitates, I quietly approach to determine whether it's an emotional issue or a language barrier'' (P9).

\subsection{Summary and Discussion of Teacher Interviews}
In this interview study, we revealed key understandings into how teachers perceive their role during in-class interventions in ITS-supported mathematics classrooms. Teachers prioritized maximizing learning, maintaining lesson momentum, sustaining student engagement through dynamic adjustments, promoting an inclusive classroom climate, and responding to individual differences. In terms of intervention types, they described a range of supports, including disciplinary guidance for error correction and scaffolding, emotional/motivational cues to boost persistence, and technical assistance for system navigation. Influencing factors included the need for equitable coverage for all students, the importance of delivering actionable cues with clear priorities, and a preference for peer-led support as a supplement to teacher intervention.
These findings are consistent with prior research on teacher orchestration in technology-enhanced classrooms. Such studies emphasize the importance of real-time cues for balancing individual and class-wide needs~\cite{alfredo_human-centred_2024}. Teachers' focus on equitable attention, such as mirrors studies of dashboard designs that reduce biases in attention allocation~\cite{an_ta_2020,holstein_co-designing_2019}. Their calls for prioritized and contextual signals build on engagement detection research that uses multimodal data to provide actionable insights~\cite{alkabbany_experimental_nodate,booth_engagement_2023,dewan_engagement_2019}.
Our interviews also highlight the importance of temporal factors, particularly help history, in building an equitable classroom for teacher intervention. Teachers stated they often used recent visits as a tie-breaker in prioritization. Existing literature on immediate states has underexplored this factor~\cite{holstein_intelligent_2017,slotta_orchestrating_nodate}.

These teacher-reported insights (especially section~\ref{factors}, where teachers said that \textit{recent help history} and \textit{current student states} guide whom they help) directly inform the subsequent quantitative analyses in Section \ref{dataandmethods}, where we empirically test whether features like help-history and student states (e.g., idle, struggling) truly predict intervention likelihood in authentic classroom data (RQ1). 
In addition, teachers reported trying to avoid repeatedly visiting the same student to support inclusion and equity, which motivates testing stickiness and directionality over time. We therefore estimate a session-lagged CLPM with both stability paths and cross-lagged paths to separate within-session co-occurrence from across-session influence (RQ2).
Because teachers aim to keep the lesson moving while supporting learning, we also ask whether visits carry forward to later mastery. We operationalize session-to-session learning using AFM-based first-mastery and test cross-lagged links between prior help and subsequent skill acquisition (RQ3), distinguishing immediate within-session gains from across-session effects.
By bridging qualitative self-reports with large-scale log analysis, we aim to validate these decision factors and uncover potential discrepancies between perceived and actual influences, informing the design of more effective analytics tools.

\section{Log Data Analysis of Teacher Helping}
\label{dataandmethods}
The new insights from the interview study motivate our focus on help history and student states in quantitative analyses. In particular, the demand for actionable cues with different priority highlights teachers' need to differentiate disengagement states like idle and struggle, guiding our use of fine-grained detectors to model these as predictors of interventions. While the prioritization of peer-led support suggests group dynamics play a role in initial help, our dataset lacks direct logs of collaboration or contextual observations (e.g., raised hands or expressions), limiting verification to teacher-logged events; thus, we concentrate on verifiable features like help history and detectable states to test temporal relationships. Notably, we make explicit the construct alignment between our interview study and the logs. Six of nine teachers had direct experience using MATHia in class; for the remaining teachers, we showed a brief Lumilo dashboard demo (Section~\ref{datacollection}) backed by an ITS with MATHia functions and learning indicators. Teachers reported that classroom math ITSs share core workflows (step-level practice, targeted feedback, mastery/progress tracking), supporting alignment between our formative constructs (Section~\ref{formativefindings}) and the log indicators.

\subsection{Dataset}
\begin{figure}
    \centering
    \includegraphics[width=\linewidth]{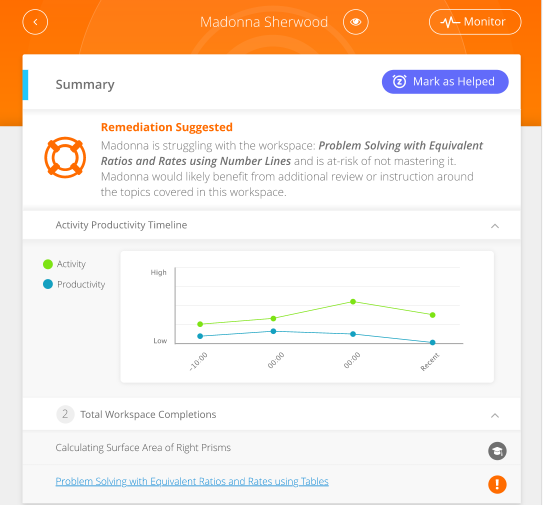}
    \caption{LiveLab dashboard shows real-time student progress monitoring and remediation suggestions for teachers. Teachers could manually log instances of providing help to students using a button ``Mark as Helped''.}
    \Description{LiveLab dashboard shows real-time student progress monitoring and remediation suggestions for teachers. A line chart visualizes trends in activity and productivity over time, while workspace completion history and current learning tasks are listed below. Teachers could manually log instances of providing help to students using a button ``Mark as Helped''.} 
    \label{fig:LiveLab}
\end{figure}

Student interaction data was collected from MATHia~\cite{fancsali2023orchestrating}, comprising 1,437,055 transactions (i.e., student actions) from 339 students enrolled in 14 middle and high school math classes across ten U.S. K-12 schools during the 2022-23 school year. These classes were selected based on the relative frequency with which teachers used a feature to mark moments of help, meaning the teacher reported through the LiveLab dashboard\footnote{MATHia is typically part of a schools’ core math curriculum and instruction in a blended model that recommends usage MATHia for 40\% of instructional time during a typical week of math instruction (e.g., two of out five instructional math periods). LiveLab is provided as optional support for orchestrating this MATHia classroom time.} (Figure~\ref{fig:LiveLab}) when they helped a specific student after 1:1 help was provided. These ``marked as helped'' instances were then recorded in the log data with timestamps, together with the student IDs, so we can relate them to students’ actions also recorded in the log data. In this classroom scenario (common to both our formative and log-based studies), students work on step-level problem solving in the ITS during regular class time while the teacher monitors progress and provides help as needed. All procedures were approved by the IRB, schools and districts authorized data use, logs were de-identified prior to analysis, and access was restricted under a data-use agreement.

\subsection{Identifying Classwork Sessions}
ITSs, including MATHia, generate transaction logs from multiple contexts, such as classwork, homework, and other in-school activities. However, teacher interventions in our study occurred during classwork sessions where instructors can provide in-the-moment support. We inferred classwork sessions from log data using the session inference algorithm reported by Gurung et al.~\cite{gurung2025starting} (see Algorithm 1). Sessions were defined based on two key criteria: (1) the presence of at least five active students, to distinguish whole-class activity from small-group or individual work, and (2) class-wide inactivity intervals, using a 7.5-minute threshold to delimit new sessions. Consistent with Gurung et al., we tested the robustness of this threshold through a sensitivity analysis, varying the inactivity interval from 2 to 30 minutes.

These identified class sessions provide the necessary underlying context for modeling the ``stickiness'' of help, that is, for modeling how teacher interventions in one session may influence teacher decisions of whom to help in subsequent sessions, independent of current cues like idle or struggle states. 
This yielded 9,565 inferred student-classwork-sessions across 14 classes,\footnote{The original dataset comprised 1.6M transactions across 15 classes. We excluded one class because it was not collected in a U.S. school, and differences in teacher intervention practices could affect the validity of the results.} with a mean session size of 15 students (SD = 5.01) and mean length of 38.29 minutes (SD = 20.31 minutes). The sessions with fewer than five active students were classified as non-classwork sessions (e.g., individual at school practice or homework sessions). The sessions outside school hours (7:00 AM - 4:00 PM) and on weekends were inferred as homework sessions and were excluded from the analysis as the teachers can only help at school. Table \ref{tab:session_stats} shows a more detailed breakdown of statistics for each of the types of sessions. Within these sessions, we aggregated teacher-logged help events, student states (e.g., idle, struggling), and interaction metrics for subsequent modeling in Sections~\ref{Detectors}.

\begin{table*}[htpb]
\centering
\caption{Session types summary. For each type, we report the number of student sessions, mean active students, mean length (minutes), and the number of student sessions with at least one logged teacher visit. Teacher help is concentrated in classwork sessions; the few ``visited'' homework/non-classwork sessions likely reflect delayed or off-class logging. These homework/non-classwork visit counts were not included in our intervention analyses.}
\label{tab:session_stats}
\begin{tabular}{lcccc}
\toprule
\textbf{Session Type} & \textbf{Number of Sessions} & \textbf{Size (students)} & \textbf{Length (min)} & \textbf{Number Visited by Teachers} \\
\midrule
Classwork & 9,565 & 15.61 & 38.29 & 364 \\
Homework & 953 & 1.15 & 21.09 & 2 \\
Non-Classwork & 2024 & 1.76 & 16.53 & 26 \\
\bottomrule
\end{tabular}
\end{table*}

\subsection{Detectors and Feature Engineering}
\label{Detectors}

We operationalized students' engagement states for each classwork session using rule-based detectors, defining two binary disengagement indicators for each session: idle (coded as 1 if the student exhibited no interactions in the software for 2 minutes or more during the session, 0 otherwise~\cite{holstein2018student}) and struggle (1 if the student attempted to master a skill three or more times without success despite system support~\cite{beck2013wheel}). These detectors drew from open-source code in prior studies on teacher augmentation tools, which have empirically shown to enhance student learning gains when integrated into dashboards for teachers in intelligent tutoring system classrooms~\cite{holstein2018student, holstein_co-designing_2019}. By capturing distinct facets of student learning behaviors, these measures are instrumental in examining teacher response patterns~\cite{fancsali2023orchestrating, karumbaiah2023spatiotemporal}.
We also defined help history as a binary feature indicating whether a given classwork session was marked as \emph{helped}. The key predictor was whether the student had prior visits or not.

\subsection{Statistical Analysis}

\subsubsection{Modeling Effects of Help History and States on Who Gets Help (RQ1)}

To rigorously test the association between prior help history and the likelihood of receiving teacher interventions, we estimated a mixed-effects logistic regression model. To adjust for contextual and behavioral covariates, we included students’ relative delayed start as a stability-sensitive indicator of motivation (prior work shows that delayed start exhibits trait-like stability across sessions and predicts outcomes beyond contemporaneous teacher behavior~\cite{gurung2025starting}), idle time (with an interaction term), and class session size as fixed effects; as student engagement is known to correlate with teacher visits and could confound the effect of prior visits \cite{karumbaiah2023spatiotemporal,jin2025who}. To account for repeated observations and clustering, we specified random intercepts for both students and classes. Models were estimated with the \texttt{glmer} function in the \texttt{lme4} package \citep{bates_fitting_2015}. We report odds ratios and confidence intervals, and additionally computed intraclass correlation coefficients (ICC) to assess variance attributable to students and classes.

\subsubsection{Modeling Directionality Across Sessions Between Help and States (RQ2)}

To test whether teacher help history is related to subsequent student states such as struggle and idle behavior, we applied a cross-lagged panel modeling (CLPM) approach to our transaction-level data. In this framework, each model simultaneously estimated (a) stability paths that capture autoregressive dependencies in student states across adjacent sessions and (b) cross-lagged paths that capture the predictive association of prior help on subsequent student states, as well as the alternative causal direction of student states predicting subsequent help. By including both directions of influence, the model directly addresses our research question of whether help history predicts later student states (causal hypothesis) or whether states primarily elicit help from teachers. 

This design satisfies key criteria for causal inference from observational data \cite{cox2018modernizing}. First, the models enforce \emph{temporal precedence} by lagging predictors one session, ensuring that the help history at time $t-1$ is used to explain states at time $t$. Second, the inclusion of \emph{outcome autocorrelation} (stability paths) measures the effect of a variable, whether it's student state or help state, on itself over time. Third, the model tests \emph{alternative causal pathways} by estimating both help $\rightarrow$ states and states $\rightarrow$ help, providing evidence for the specificity and directionality of effects. Finally, the magnitude and significance of the cross-lagged coefficients indicate the degree of the predictive associations between constructs controlled for the autoregressive effects of individual variables, allowing us to assess the robustness of associations across students and classes. 

Together, these features make CLPM a better choice of a causal modeling strategy for RQ2 than bivariate tests or single-equation regressions: it allows us to examine whether teacher interventions have a lasting and prospective influence on subsequent student states, or whether teacher attention is better explained as a response to those states.

\begin{figure}[htpb]
\centering
\resizebox{\columnwidth}{!}{%
\begin{tikzpicture}[
    node distance=2.5cm and 3.0cm,
    latent/.style={draw, rectangle, minimum height=1.2cm, minimum width=2.4cm, align=center, thick},
    arrow/.style={->, thick},
    covar/.style={<->, thick, gray, opacity=0.7},
    stability/.style={->, thick, blue},
    crosslagged/.style={->, thick, red}
]

\node[latent] (h1) {Helped Session\\$h_1$};
\node[latent, right=of h1] (h2) {Helped Session\\$h_2$};
\node[latent, below=of h1] (s1) {Idle/Struggle\\$s_1$};
\node[latent, below=of h2] (s2) {Idle/Struggle\\$s_2$};

\draw[stability] (h1) -- node[above] {$\beta_{h}$} (h2);
\draw[stability] (s1) -- node[below] {$\beta_{s}$} (s2);

\draw[crosslagged] (h1) -- node[sloped, above, pos=0.35] {$\lambda_{hs}$} (s2);
\draw[crosslagged] (s1) -- node[sloped, above, pos=0.35] {$\lambda_{sh}$} (h2);

\draw[covar] (h1) to[bend right=30] node[left] {\scriptsize$\text{Cov}(h_1, s_1)$} (s1);
\draw[covar] (h2) to[bend left=30] node[right] {\scriptsize$\text{Cov}(h_2, s_2)$} (s2);

\end{tikzpicture}
}
\caption{Cross-lagged panel model with two time points illustrating the relationship between teacher help history ($h$) and student states such as idle or struggle ($s$). Stability effects ($\beta_h, \beta_s$) are shown in blue, cross-lagged effects ($\lambda_{hs}, \lambda_{sh}$) in red, and covariances are in gray.}
\Description{Cross-lagged panel model with two time points illustrating the relationship between teacher help history ($h$) and student states such as idle or struggle ($s$). Stability effects ($\beta_h, \beta_s$) are shown in blue, cross-lagged effects ($\lambda_{hs}, \lambda_{sh}$) in red, and covariances are in gray.}
\label{fig:cross_lagged_model}
\end{figure}
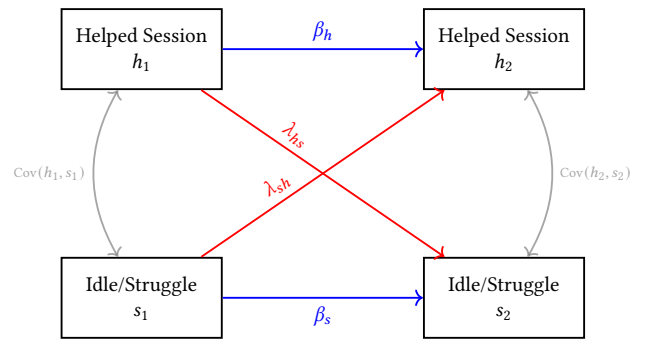

\subsubsection{Investigating Relation Between Teacher Help and Later Learning (RQ3)}

To investigate whether teacher interventions contribute to subsequent student learning, we focused on the acquisition of new knowledge components (KCs), operationalized as the first occasion when a student’s predicted mastery of a KC exceeded an 80\% threshold. This mastery estimate was derived from an additive factors model (AFM) trained on student problem-solving attempts, providing a session-level indicator of whether the student had acquired a new skill. 

To operationalize students’ learning progress in MATHia, we estimated an Additive Factors Model (AFM), a standard logistic regression model widely used in knowledge tracing research \cite{cen2006learning}. The AFM assumes that the probability of a student answering a step correctly increases as a linear function of the number of prior opportunities to practice the corresponding knowledge component (KC). The model includes random intercepts for both students and KCs, thereby accounting for baseline differences in student ability and skill difficulty.

Formally, the AFM model is specified as:
\[
\text{logit}(P(\text{correct}_{i,j})) = \theta_j + \beta_k + \gamma \cdot \text{opportunity}_{i,jk},
\]
where $\theta_j$ is the intercept for student $j$, $\beta_k$ is the intercept for skill $k$, and $\gamma$ is the fixed learning rate associated with each additional practice opportunity. The key assumption is that learning accrues additively and linearly with practice, such that each opportunity provides an equal marginal gain in log-odds of correctness.

We then used the fitted AFM model to generate predicted mastery probabilities for each student–KC–opportunity instance. A KC was considered \emph{newly acquired} once the predicted probability of correctness exceeded an 80\% threshold for the first time. This thresholding procedure provided session-level indicators of skill acquisition, which served as the dependent variable in our cross-lagged panel analyses of teacher help and subsequent learning.

We then applied a CLPM framework (similar to RQ2) to jointly estimate the dynamics between teacher help and new KC acquisition across adjacent sessions. In this model, each outcome (teacher help and new KC acquisition) was regressed on its own lagged value to capture stability over time and on the lagged value of the other construct to capture potential cross-lagged influences. Specifically, we tested whether receiving help in session $t-1$ predicted the number of new KCs acquired in session $t$, and conversely, whether acquiring new KCs predicted whether the student was helped in the following session. As in RQ2, we included covariances between concurrent residuals as well as among lagged predictors, thereby adjusting for baseline differences in student trajectories and controlling for within-session associations between help and learning.

This design allows us to distinguish between short-term correlations (e.g., students who receive help also appear to learn more within the same session) and longitudinal effects (e.g., whether help in one session predicts increased learning opportunities in future sessions). By explicitly modeling temporal precedence and alternative pathways, the CLPM framework provides a more rigorous test of whether teacher visits prospectively support students’ learning progress, as opposed to being a reactive signal that identifies students who are already struggling or thriving.

\section{Results}

\begin{figure*}[htbp]
  \centering
  \begin{minipage}[t]{\columnwidth}
    \centering
    \includegraphics[width=\linewidth]{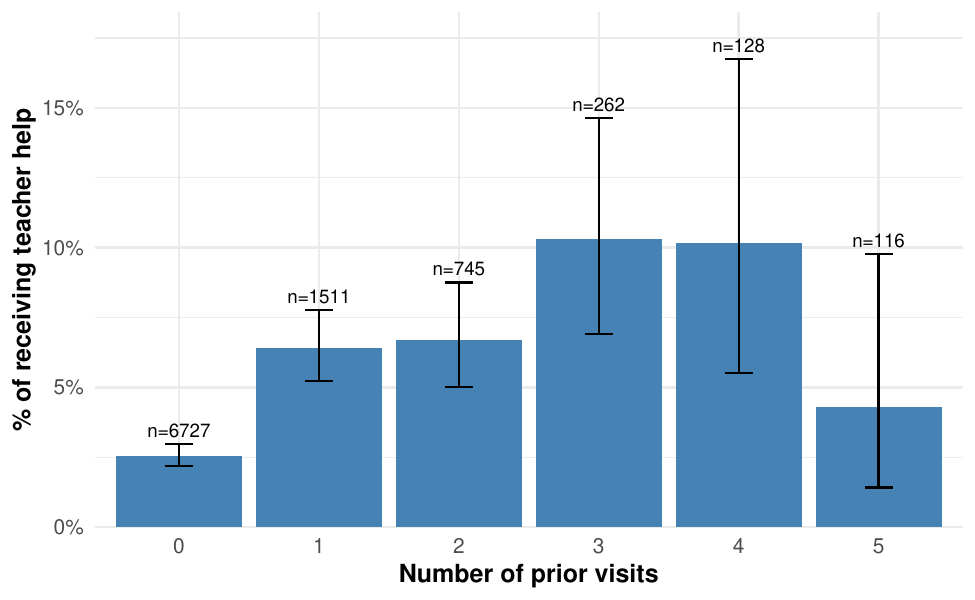}
    \Description{Proportion of student sessions with teacher help by number of prior visits (0--5). Error bars show 95\% exact binomial confidence intervals.}
    \captionof{figure}{Proportion of student sessions with teacher help by number of prior visits (0--5). Error bars show 95\% exact binomial confidence intervals.}
    \label{fig:help_prior_visits}
  \end{minipage}\hfill
  \begin{minipage}[t]{\columnwidth}
    \centering
    \includegraphics[width=0.75\linewidth]{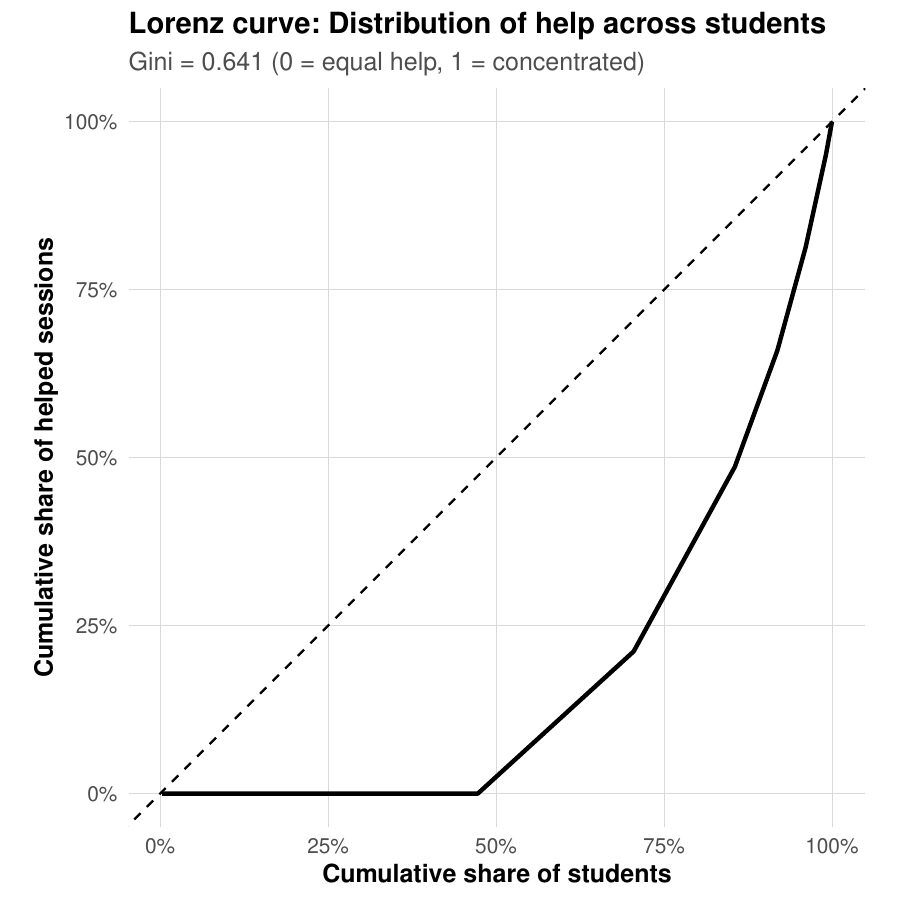}
    \Description{Lorenz-style concentration curve of teacher help across student sessions. A stronger departure from the $45^{\circ}$ line indicates greater inequality.}

    \captionof{figure}{Lorenz-style concentration curve of teacher help across student sessions. A stronger departure from the $45^{\circ}$ line indicates greater inequality.}
    \label{fig:lorenz_help}
  \end{minipage}
\end{figure*}

Descriptively, we find that teacher interventions were more common for students who had previously received help. Among students without a prior help history, only 2.6\% of their classwork sessions were marked as helped. In contrast, for students who had been helped before, the proportion rose to 6.8\%. This association was statistically significant, $\chi^2(1, N{=}9{,}565) = 95.42, p < .001$, indicating that students with a history of prior help were more likely to receive additional teacher interventions. Figure~\ref{fig:help_prior_visits} further illustrates this trend: the likelihood of receiving help increased steadily across one to three prior visits, reaching around 10\% of sessions, compared to only 3\% when no prior visits were recorded. While the confidence intervals are wide for students with four or more prior visits (reflecting smaller sample sizes), the overall pattern suggests that teacher attention tends to recur and accumulate for students once they have entered a help-receiving trajectory.

Before turning to model-based analyses, we quantify how teacher help is distributed across students to ensure sufficient variability for studying persistence. Figure~\ref{fig:lorenz_help} plots a Lorenz-style concentration curve: the cumulative share of \emph{helped sessions} on the $y$-axis against the cumulative share of students ranked by their total number of helped sessions on the $x$-axis, with the dashed $45^{\circ}$ line indicating perfect equality \citep{gastwirth1972estimation}. The curve departs substantially from the equality line, indicating high inequality in the distribution of teacher help across students (i.e., a relatively small subset of students accounts for a disproportionate share of visits). In particular, in the average session, about 50\% of students do not get visited, while the other 50\% receive all the help. This concentration supports our later analyses of persistence by showing that help is not evenly spread but clustered and, potentially, sticks to specific students over time.

\subsection{RQ1: Effects of Help History and States on Who Gets Help}

Results from the mixed-effects logistic regression model confirmed the descriptive pattern. Students with a prior history of receiving help were significantly more likely to receive additional teacher interventions ($\mathrm{OR} = 1.32$, 95\% CI [1.19, 1.46], $p < .001$), even after adjusting for student disengagement and class-level differences. Beyond prior visits, we also found that idle time was a strong positive predictor of help ($\mathrm{OR} = 2.74$, 95\% CI [1.83, 4.11], $p < .001$), and that larger class session sizes were modestly but significantly associated with higher odds of intervention ($\mathrm{OR} = 1.08$, 95\% CI [1.05, 1.11], $p < .001$). Relative delayed start was not a significant predictor of teacher help ($\mathrm{OR} = 0.89$, 95\% CI [0.63, 1.24], $p = .483$), and its interaction with idle time was likewise non-significant ($\mathrm{OR} = 1.25$, 95\% CI [0.87, 1.79], $p = .222$).

The random effects structure indicated meaningful clustering: approximately 16\% of the variance in help interventions was attributable to differences between students and classes (ICC = .16). Model fit statistics suggested that fixed effects explained about 11\% of the variance in teacher interventions ($R^{2}_{\mathrm{marginal}} = .107$), while the full model including random effects explained 25\% ($R^{2}_{\mathrm{conditional}} = .250$). These findings indicate that teacher interventions are more likely to recur for students with a help history and are also shaped by contextual engagement factors such as idle time and class size.

\subsection{RQ2: Directionality Across Sessions Between Help and States}

Both autoregressive effects were significant, indicating stability across sessions: prior help strongly predicted subsequent help ($a1 = 0.37$, $z = 8.24$, $p < .001$), and prior idle behavior also predicted later idle ($c1 = 0.12$, $z = 5.21$, $p < .001$). These coefficients imply that both teacher intervention patterns and idle states are not random, but tend to persist once established—students who were helped in one session are more likely to be helped again, and those who were idle once are somewhat more likely to idle again. 

Turning to the cross-lagged paths, we find an asymmetric pattern. Prior idle behavior significantly reduced the likelihood of subsequent teacher help ($b1 = -0.10$, $z = -2.62$, $p = .009$), suggesting that when students were idle in the preceding session, they were about 10\% less likely to receive teacher help in the next session. In contrast, prior help predicting subsequent idle was not statistically significant ($d1 = -0.05$, $z = -1.45$, $p = .147$), indicating no clear evidence that receiving help changed the likelihood of students becoming idle later. A Wald test confirmed that the joint set of cross-lagged paths was significant, $\chi^2(2) = 8.96$, $p = .011$, but the two paths were not significantly different in magnitude ($\chi^2(1) = 0.82$, $p = .365$). 

Taken together, these results provide little evidence for a causal hypothesis that teacher interventions directly alter subsequent idle behavior. Instead, the direction of influence is more consistent with the reverse: prior idle states predict reduced teacher intervention, even after adjusting for the stability of both constructs. In practice, this suggests that teachers are more consistent in revisiting students they already helped than in responding to transient idle states. The persistence of help may reflect teachers’ tendency to monitor the same students over time, while the lack of effect of help on idle implies that teacher interventions, at least as observed here, do not substantially shift students’ long-term engagement states.

\subsection{RQ3: Relation Between Teacher Help and Later Learning}

The Additive Factors Model (AFM) fit to 425{,}285 first attempts (commonly filtered for AFM estimation, see \cite{koedinger2010data}) across all students and 827 KCs confirmed a robust learning effect: each additional practice opportunity increased the odds of a correct response by about 0.9\% ($\gamma = 0.0092$, $SE = 0.00028$, $z = 32.78$, $p < .001$). Variance components indicated substantial heterogeneity at both the student ($\sigma^2 = 0.29$) and skill ($\sigma^2 = 1.26$) levels.

Using predicted mastery probabilities from the AFM, we flagged a skill as newly acquired once a student’s estimated probability of correct performance first exceeded 80\%. Cross-lagged panel analyses of help and new skill acquisition showed a clear asymmetry. Teacher help in session $t-1$ did not significantly predict the number of new skills learned in session $t$ ($b_K = 0.04$, $z = 0.64$, $p = .52$). By contrast, prior skill acquisition significantly lowered the likelihood of receiving help in the subsequent session ($b_H = -0.04$, $z = -5.39$, $p < .001$). This suggests that teachers actively redirect attention away from students demonstrating progress. Within sessions, help and learning were positively correlated ($r \approx .13$, $p < .001$), consistent with teachers intervening in ways that support immediate learning, but these effects did not persist into future sessions, as indicated by the cross-lagged effects.

Overall, these findings indicate that teacher help is reactive and short-term: it co-occurs with learning within a session but does not translate into additional skill acquisition in later sessions. Instead, teachers appear to allocate their limited time by concentrating future visits on students who have not recently shown evidence of new learning.

\section{Discussion}

Our findings from teacher interviews and quantitative analyses of log data highlight the temporal complexities of teacher interventions in ITS-supported mathematics classrooms. Teachers reported prioritizing equitable attention and immediate student states, yet empirical models revealed persistent ``sticky'' help for previously supported students and only fleeting effects of these help interventions on disengagement and learning. These discrepancies inform both practice and design in learning analytics, as we discuss next.

\subsection{Research Implications for Teacher Orchestration and Learning Analytics}

On first glance, our GLMM results for RQ1 are somewhat counterintuitive: students who were helped once remained more likely to be helped again, whereas students who idled once were \textit{less likely} to be helped. 
This finding contrasts with teachers’ self-reported intentions in the interview study, where they emphasized visiting every student and distributing attention fairly. Importantly, teachers also acknowledged that students differ in need, and that ``fairness'' does not necessarily mean spending equal time with every student; rather, it often involves balancing responsiveness to need with avoiding over-focusing on a small subset of students. Our log-based analyses speak to the patterning of visits over time, but they do not directly test whether teachers’ allocation matches students’ needs under any particular definition of fairness.
The results suggest that help history functions as an anchor: students who were helped once are seen as warranting continued checks, while idle is treated as a momentary state that prompts a short, reactive response. The cross-lagged model makes this distinction explicit. Within a session, idle and help co-occur because teachers step in when disengagement appears. Across sessions, after accounting for this immediate co-occurrence, idle alone does not predict continued monitoring. In practice, teachers seem to treat idleness as temporary, whereas a recorded help event marks a student for follow-up. %
At the same time, multiple alternative explanations are plausible. Persistence of visits could reflect stable differences in student need (e.g., some students consistently struggle more), stable differences in help-seeking behavior (e.g., some students more often request help), or classroom norms and grouping structures that repeatedly place teachers near the same students. Thus, while the data are consistent with the interpretation that teachers engage in ongoing follow-up once support has been initiated, we cannot distinguish proactive monitoring from reactive revisits based on logs alone, and we cannot infer whether the resulting allocation would be judged ``fair'' by teachers themselves. Future work could triangulate logs with observations~\cite {karumbaiah2023spatiotemporal} and elicit teachers’ interpretations by showing them visualizations of their own visit distributions over time to understand how they define equitable allocation and how they would adjust practice in response.

If the goal of teacher intervention is to counter sticky help and surface inequities, work in equity-forward analytics suggests surfacing attention gaps and prompting concrete next steps, not only risk cues~\cite{holstein_co-designing_2019, sloan2024equity}. Prior work notes that, despite growing attention to equity in education, learning analytics has only begun to address this area in depth~\cite{uttamchandani2022introduction}. Our results add evidence of ``sticky help,'' which may worsen equity by concentrating attention on the same students. Future research should pair log analyses with field observations to detect systematic under-service of specific students or groups, and calibrate dashboards. For example, adding coverage-aware features (e.g., a panel showing under-visited students; soft caps or cooldowns on repeated visits; a start-of-session ``sweep'' task) so teachers can deliberately offset history-anchored patterns. Studies of handheld and classroom dashboards show teachers value tools that make orchestration doable—bridging the gap between signals and next actions~\cite{martinez-maldonado_handheld_2019}. Future work should also focus on measuring attention distribution and learning equity across the session. It could also examine teacher sensemaking supports that translate analytics into planned classroom moves~\cite{kitto2017designing}. 

For RQ3, we found that within the same session, help and learning are positively correlated. Across sessions, prior help does not predict more new skills next time, while recent new-skill acquisition predicts less help next time. This echoes the help-seeking literature in ITSs, where help ``helps, but only so much,'' with benefits strongest in the moment and confounded by who seeks/receives help~\cite{aleven2016help}. 
We do not take this to mean that teacher visits are ineffective or that additional visits could not help. Rather, several mechanisms can explain the observed pattern. First, this pattern may be that selection into help likely remains. Even with controls, teachers tend to target students with greater need, who have lower expected gains on the next session, which can attenuate estimated prospective effects of help. Also, the main pathway of help may be short-lived. In-class support often resolves an immediate hurdle without shifting strategies or motivation enough to raise mastery in the next session, where students are likely working on different content (e.g., different problem types and skills). Another possible interpretation is that some teachers help functions like answer-giving or very short worked examples, raising performance in the moment without the deeper scaffolding needed for lasting learning (a pattern also debated in language-model–assisted tasks: higher short-term performance, weaker durable learning~\cite{bastani2024generative}). Future work should closely analyze what actually happens during help, record and code help episodes (e.g., answer-giving vs prompting vs explanation), link help types to next-session outcomes, and experimentally compare brief answer-like support to scaffolded prompts that require student planning, monitoring, and self-explanation. Third, session-to-session estimates may be diluted by unobserved context (e.g., unlogged brief interventions, peer support, teacher style differences) that we discuss in our limitations. Future work should collect richer context (e.g., brief in-class help logs, peer-support flags, teacher annotations), integrate multimodal signals where feasible (e.g., time-aligned off-software interactions from classroom audio/video or multimodal analytics dashboards~\cite{dewan_engagement_2019,echeverria2024teamslides}), and test session-aware interventions in randomized or quasi-experimental designs to separate selection from effect.

\subsection{Limitations and Future Work}
One limitation is our reliance on teacher‐logged help events in LiveLab, which may miss brief or unrecorded interventions. %
Because our classes were selected based on relatively frequent use of this logging feature, the sample is not randomly drawn and may overrepresent teachers who are more consistent or data-oriented in documenting visits, limiting generalizability to all ITS classrooms. Completeness is also uncertain (across all inferred classwork sessions, only around 4\% contained a teacher-logged help event), and any missingness could bias estimates of ``sticky help'' and state-response in unknown directions (e.g., short visits may be less likely to be logged, but longer visits may also reduce opportunities to log). In our models, help history is operationalized as a binary indicator (prior visit vs.\ not); future work should test count-based specifications (e.g., number of prior visits, within-session revisit counts) and, where possible, incorporate visit duration or proximity-based sensing to better characterize support intensity.
Future work should therefore verify log completeness with classroom observations or automated logging (e.g., time-synchronized teacher positioning to align proximity-based visits with tutor logs)~\cite{karumbaiah2023spatiotemporal}, and collect teacher/class covariates (including cross-level interactions) to isolate practice differences.

Additionally, we lack data on offline behaviors and collaborative activity, which our teacher interviews identified as important factors in intervention decisions. Without logs of peer support or non-digital cues, we cannot fully connect decisions to these contexts noted in interviews. This likely does not overturn help persistence (offline disengagement may track with logged idle), but may understate how group dynamics mitigate idle. We excluded visits labeled as homework or non-classwork; some may map to sessions, but without context, we could not verify them. Future work should add multimodal data (e.g., time-aligned audio/video) to capture offline actions and test whether they moderate the fleeting nature of idle. 
Furthermore, the dataset does not record the type/content of interventions. This prevents analyzing whether ``sticky help'' is driven by persistent instructional needs or by recurring motivational regulation, and motivates future work that record, codes or infers intervention types using classroom observations or richer logging.

Our dataset for quantitative analysis does not include detailed instructor information (e.g., years of experience, training with ITSs, instructional style), which limits generalization across teacher profiles. Prior work on demographic match and teacher expectations shows mixed evidence~\cite{harbatkin2021does, redding2019teacher}, suggesting these factors might confound who gets help and how often. While focusing on MATHia in U.S. middle schools provides ecological validity, findings may not extend to other systems, grades, or cultural contexts. In addition, our quantitative analyses are restricted to MATHia logs. Expanding to multiple subjects and platforms, and modeling platform/subject moderation, would help distinguish general patterns from system-specific effects. %

Finally, because this study is observational, the cross-lagged models do not establish causal effects; rather, they provide evidence about the relative plausibility of alternative directional hypotheses. Future work should expand to diverse settings, include teacher demographics as moderators, and test causal pathways in experiments. 

\section{Conclusion}
We contribute a novel investigation of the temporal dynamics of teacher interventions across multiple sessions in technology-supported mathematics classrooms. Drawing on teacher interviews and large-scale log data analyses, our findings indicate that prior help history predicts subsequent intervention decisions, independent of current student states such as idleness or struggle. Cross-lagged panel models further reveal that help is ``sticky,'' recurring across sessions for previously supported students, whereas effects on learning and engagement are fleeting, since help did not predict sustained benefits on learning.
These results sit in tension with teachers’ stated aspirations to cover the class, but whether the observed revisit pattern is problematic depends on how equity is defined (e.g., equal time versus time relative to need). This highlights the importance of designing real-time orchestration tools that make attention patterns visible and support deliberate trade-offs.

Our contributions include empirical validation of help history as a key decision factor, novel application of temporal modeling to authentic classroom data, and insights into the short-lived nature of help effects on engagement and learning. For learning analytics, these results motivate dashboards that summarize visit distributions and revisits over time to support teacher reflection and intentional allocation decisions under teachers’ own definitions of fairness and student need, rather than presuming that ``sticky''  patterns are necessarily inequitable.
Future work could extend these analyses by designing experiments to test causal mechanisms, incorporating multimodal data on offline behaviors, and exploring generalizability across diverse educational contexts. Ultimately, by addressing these temporal imbalances, learning analytics can better support teachers in fostering inclusive and effective learning environments.

\begin{acks}
This research was funded by the Institute of Education Sciences (IES) of the U.S. Department of Education (Award \#R305A240281). We also thank all participating teachers for their time and contributions to this study.

\end{acks}

\bibliographystyle{ACM-Reference-Format}
\bibliography{main}

\appendix

\end{document}